\documentclass[aps,pra,floatfix,amsmath,preprint,12pt]{revtex4}
\usepackage{graphicx}
\usepackage{amsthm}
\usepackage{relsize}
\usepackage{amsmath,amssymb}
\usepackage{amsfonts}
\usepackage{color}
\usepackage{xcolor}
\usepackage{array}
\usepackage{ulem}
\usepackage{float}
\restylefloat{figure}
\begin{document}

\title{A set theoretical approach for the partial tracing operation in quantum mechanics}

\author{Pranay Barkataki}
\email{pranaybarkataki@students.vnit.ac.in}
\author{M. S. Ramkarthik}
\email{msramkarthik@phy.vnit.ac.in}
\affiliation{Department of Physics, Visvesvaraya National Institute of Technology, Nagpur - 440010}
\date{\today}

\vspace{1cm}

\begin{abstract}
Partial trace is a very important mathematical operation in quantum mechanics. It is not only helpful in studying the subsystems of a composite quantum system but also used in computing a vast majority of quantum entanglement measures. Calculating partial trace becomes computationally very intensive with increasing number of qubits as the Hilbert space dimension increases exponentially. In this paper we discuss about our new method of partial tracing which is based on set theory and it is more efficient. The proposed method of partial tracing overcomes all the limitations of the other well known methods such as being computationally intensive and being limited to low dimensional Hilbert spaces. We give a detailed theoretical description of our method and also provide an explicit example of the computation. The merits of the new method and other key ideas are discussed.
\end{abstract}

\maketitle

\newcommand{\newc}{\newcommand}
\newc{\beq}{\begin{equation}}
\newc{\eeq}{\end{equation}}
\newc{\kt}{\rangle}
\newc{\br}{\langle}
\newc{\beqa}{\begin{eqnarray}}
\newc{\eeqa}{\end{eqnarray}}
\newc{\pr}{\prime}
\newc{\longra}{\longrightarrow}
\newc{\ot}{\otimes}
\newc{\rarrow}{\rightarrow}
\newc{\h}{\hat}
\newc{\bom}{\boldmath}
\newc{\btd}{\bigtriangledown}
\newc{\al}{\alpha}
\newc{\be}{\beta}
\newc{\ld}{\lambda}
\newc{\sg}{\sigma}
\newc{\p}{\psi}
\newc{\eps}{\epsilon}
\newc{\om}{\omega}
\newc{\mb}{\mbox}
\newc{\tm}{\times}
\newc{\hu}{\hat{u}}
\newc{\hv}{\hat{v}}

\section{Introduction}
In quantum mechanics a general state of the quantum system can be represented by a density matrix $\rho$ \cite{n.c.} from which we can calculate 
every observable quantity related to the system. A composite quantum system which is composed of two subsystems is described by the density matrix 
$\rho^{_{AB}}$, where $A$ and $B$ 
represent the subsystems. Given the density matrix of the composite system, if we have to determine the quantum state of the subsystems, then the mathematical
operation we perform is a map called the partial trace \cite{n.c.}. This map acts on the density matrix $\rho^{_{AB}}$ of the total system and as a result gives
us the states of the individual subsystems which are in turn described by the corresponding reduced density matrices \cite{peres}. Mathematically the reduced density matrix of subsystem
$A$ is written as $\rho^{_{A}}=tr_{_{B}}(\rho^{_{AB}})$, what we have done is to take the trace over the complimentary subsystem. The partial trace
is the {\it only} operation which is used to determine the states of the subsystems. Not only this, the reduced density matrices and their eigenvalues
are the most important ingredients in computing a vast majority of the entanglement measures \cite{ent} used in quantum information theory and other allied 
areas of physics. The entanglement of a system is measured by von Neumann entropy \cite{neu_ent} for two-qubit pure states, block entropy \cite{b.ent} for multipartite pure states and 
concurrence \cite{conc} for a two-qubit pure or mixed state. It is also well known that if we have a
composite system to be an entangled state, then partially tracing out a subsystem results in the reduced density matrix being a mixed state and this is one of the central features of the phenomenon of entanglement. Due to the aforesaid facts, partial tracing becomes an important mathematical operation in
quantum mechanics as a whole.
An efficient way to calculate the partial trace of any quantum state has been recently reported \cite{jonas} and we will be discussing this in section \ref{jmpt} for completeness. However, our new method which will be discussed at length in section \ref{cmpt} of this paper for calculating partial trace is much more simpler, effective and has other interesting features. Presently, with the thin line between quantum information theory and condensed matter physics disappearing in the last few years, our method can be of value to such condensed matter physicists who work on many body problems and finite systems \cite{b.e.}.
\section{ The Usual method to calculate partial trace}
If we a have density matrix $\rho^{_{AB}}$ and we partially trace off subsystem $B$ out of the density matrix $\rho^{_{AB}}$, the mathematical operation
can be written as :
\begin{equation}
 \rho^{_{A}}=tr_{_{B}}(\rho^{_{AB}})=\sum_{j=1}^{\eta_{_{B}}}(I_a\otimes \langle b_j\vert)\rho^{_{AB}}(I_a\otimes \vert b_j\rangle), \label{rdm1}
\end{equation}
where $\eta_{_{B}}$ is the dimension of the Hilbert space and $\vert b_j\rangle$'s are the basis vectors of the subsystem $B$. The operator $I_a$ is the 
identity matrix in the Hilbert space of the subsystem $A$. If we have multiple subsystems say $A,B,C$ of Hilbert space dimension $\eta_{_{A}},
\eta_{_{B}},\eta_{_{C}}$ respectively, we can trace out two subsystems $B,C$ out of the density matrix $\rho^{_{ABC}}$, then the reduced density matrix $\rho^{_{A}}$ is written as :
\begin{eqnarray}
 \rho^{_{A}}&=&tr_{_{BC}}(\rho^{_{ABC}})=tr_{_{B}}\{tr_{_{C}}(\rho^{_{ABC}})\}=tr_{_{C}}\{tr_{_{B}}(\rho^{_{ABC}})\},\\
 &=&\sum_{i=1}^{\eta_{_{B}}}\sum_{j=1}^{\eta_{_{C}}}(I_a\otimes \langle b_i\vert\otimes\langle c_j\vert)\rho^{_{ABC}}(I_a\otimes\vert b_i\rangle\otimes
 \vert c_j\rangle).
\end{eqnarray}
If the dimension of the Hilbert space of the composite system is large then partially tracing out subsystems becomes more and more complex as the number 
of matrix multiplications and the number of associated inner products to be calculated increases exponentially. This makes the computation of the reduced density matrix difficult both analytically and numerically. 
\section{calculating partial trace } \label{jmpt}
\subsection{Bipartition system}
We a have density matrix $\rho$ of a composite system whose subsystems are $A$ and $B$ with Hilbert space dimension $\eta_{_{A}}$ and $\eta_{_{B}}$ respectively. In the 
density matrix $\rho$ we partially trace out any one of the subsystem, let us consider we partially trace out subsystem $B$. We had devised a method 
independently but later it was brought to our notice that it was already published {\cite{jonas}}. For partially tracing out any one of the subsystem of a bipartition system this method is much more faster than the usual method of partial tracing. The state vector $\vert j\rangle$ spans the Hilbert space of the subsystem $B$. It has only one non-null element in the $j^{th}$ position of the matrix of dimension 
$\eta_{_{B}}\times1$, so when we calculate the reduced density matrix $\rho^{{A}}$ by tracing out the subsystem $B$ which now spanned in the 
$\vert j\rangle$ basis, we do not multiply with null elements of $\vert j\rangle$ but simply add those terms which
will contribute to the matrix element of the resulting reduced density matrix. On replacing $\vert b_j\rangle$ by $\vert j \rangle$ in Eq.[\ref{rdm1}] it can be
trivially seen that the matrix element $\rho^{{A}}_{k,l}$ is
\begin{equation}
 \rho^{{A}}_{k,l}=\sum_{j=1}^{\eta_{_{B}}}\rho_{(k-1)\eta_{_{B}}+j,(l-1)\eta_{_{B}}+j}.
 \label{jmpt_bp}
\end{equation}
\subsection{Multiple partition system}
In the paper \cite{jonas}, the author has given a method for computing 
partial trace of multiple partitions system which is faster than usual method of calculating partial trace. We give a birds eye view of this method. The composite system density matrix $\rho^{ABC}$ 
has three subsystems $A,B,C$ of Hilbert space dimensions $\eta_{_{A}}$, $\eta_{_{B}}$, $\eta_{_{C}}$ respectively.
We write the matrix $\rho^{ABC}$ in the computational basis as follows
\begin{equation}
 \rho^{ABC}=\sum_{j,m=1}^{\eta_{_{A}}}\sum_{k,n=1}^{\eta_{_{B}}}\sum_{l,o=1}^{\eta_{_{C}}} \langle jkl\vert\rho^{ABC}\vert mno\rangle (\vert j\rangle\langle
 m\vert)\otimes(\vert k\rangle\langle n\vert)\otimes(\vert l\rangle\langle o \vert).
\end{equation}
Partially tracing out subsystem $B$ from the composite system $ABC$ to get the reduced density matrix $\rho^{AC}$ as under
\begin{eqnarray}
 \rho^{_{AC}}&=&\sum_{j,m=1}^{\eta_{_{A}}}\sum_{k,n=1}^{\eta_{_{B}}}\sum_{l,o=1}^{\eta_{_{C}}} \langle jkl\vert\rho^{ABC}\vert mno\rangle (\vert j\rangle\langle
 m\vert)\otimes tr_{_{B}}(\vert k\rangle\langle n\vert)\otimes(\vert l\rangle\langle o \vert),\label{rdm_abc}\\
 \rho^{_{AC}}&=&\sum_{j,m=1}^{\eta_{_{A}}}\sum_{l,o=1}^{\eta_{_{C}}}\bigg(\sum_{k=1}^{\eta_{_{B}}}\langle jkl\vert\rho^{ABC}\vert mko\rangle\bigg) (\vert j\rangle\langle
 m\vert)\otimes(\vert l\rangle\langle o \vert),\\
 \rho^{_{AC}}&=&\sum_{j,m=1}^{\eta_{_{A}}}\sum_{l,o=1}^{\eta_{_{C}}}\langle jl\vert\rho^{AC}\vert mo\rangle(\vert j\rangle\langle
 m\vert)\otimes(\vert l\rangle\langle o \vert) \label{rdm_ac}.
\end{eqnarray}
The state vector $\vert mno\rangle$ has only one non null element in the matrix at the position $\alpha=(m-1)\eta_{_{B}}\eta_{_{C}}+(n-1)\eta_{_{C}}
+o$, the matrix $\rho^{ABC}\vert mno\rangle$ is equivalent to $\alpha^{th}$ column of the matrix $\rho^{ABC}$. Using this fact in Eq.[\ref{rdm_abc}] and Eq.[\ref{rdm_ac}] results in the following after some calculations, 
\begin{equation}
 {\rho}^{_{AC}}_{{(j-1)\eta_{_{C}}+l,(m-1)\eta_{_{C}}+o}}=\sum_{k=1}^{\eta_{_{B}}}\rho_{{(j-1)\eta_{_{B}}\eta_{_{C}}+(k-1)\eta_{_{C}}+l,(m-1)\eta_{_{B}}\eta_{_{C}}+(k-1)\eta_{_{C}}+o}}.
\label{jmpt_mp} 
\end{equation}
As it can be easily seen, there are no inner products or matrix multiplications to be performed. We can simply generate the reduced density matrix by adding the terms of the composite system density matrix that contributes to the matrix element of the corresponding reduced density matrix of the subsystem.

\section{Theoretical framework of our method to calculate Partial Trace of Multiple Partitions using set theory} \label{cmpt}
We have seen in the previous sections that the method of calculating partial trace, becomes cumbersome as the number of qubits increases. It is also to be noted that the method as mentioned in \cite{jonas} is quite nontrivial when used in the calculation of the reduced density matrices of any random partitions (like for example, partial tracing all odd or even spins out of a collection of spins). Keeping these requirements in mind, we have devised a new method which we present in this section. This method considerably simplifies the calculation of partial tracing both analytically and numerically. In this new method of partial tracing we consider a quantum system of $N$ qubits which is represented by the density matrix $\rho$ whose matrix elements are labeled from $0$ to $2^{N}-1$. We divide the system into $N$ subsystems and each of the subsystems has a Hilbert space of dimension two. We trace out $M$ qubits $(M< N)$ out of the quantum system and the position of the traced out qubits is represented as $\{i_k, \forall\ k=1,2,\cdots,M\}$. The binary place value of the traced out qubits is written as
\begin{equation}
\lambda_{1}=2^{i_1-1};\lambda_{2}=2^{i_2-1};\cdots;\lambda_{M}=2^{i_M-1}
\label{lam}
\end{equation}
We calculate the matrix element $\rho^\prime_{l,m}$ of the reduced density matrix $\rho^\prime$ $(2^{N-M}\times2^{N-M})$. First we  have to convert the decimal numbers $l$ and $m$ into binary equivalent numbers represented by $p$ and $q$ respectively. The binary numbers $p$ and $q$ are a string of $N-M$ numbers of $0's$ and $1's$, we take another set of binary numbers $p^{\prime}$ and $q^{\prime}$ with string of $N$ numbers. In $p^{\prime}$ the $i^{th}_1,i^{th}_2,\cdots, i^{th}_M$ positions in the string of $N$ numbers will be replaced by $0$ and in the rest of the positions of the string will be replaced sequentially by the $N-M$ numbers of the binary number $p$ and a similar procedure is followed for the number $q^{\prime}$. The decimal equivalent of the binary numbers $p^{\prime}$ and $q^{\prime}$ are $l^{\prime}_1$ and $m^{\prime}_1$. Now we define a set $S$ whose elements are the binary place values as shown below
\begin{equation}
S=\{\lambda_i,\forall\ i=1,2,\cdots,M\}.
\end{equation}
The formalism which we have developed above lends itself for a natural generalization via the concept of power set \cite{set} in abstract algebra. $P(S)$ is defined as the power set of the set $S$ and it can be written as
\begin{equation}
P(S)=\bigg\{\big\{\big\},\big\{\lambda_1\big\},\cdots,\big\{\lambda_M\big\},\big\{\lambda_1,\lambda_2\big\},\cdots,\big\{\lambda_1,\lambda_2,\lambda_3\big\},\cdots,\big\{\lambda_1,\lambda_2,\cdots,\lambda_M\big\}\bigg\}
\label{PS}
\end{equation}
The number of elements in the power set $P(S)$ is written as
\begin{equation}
\vert P(S)\vert=2^{\vert S\vert}=2^M.
\label{PSE}
\end{equation}
The elements in the power set can be represented by $\eta_k$ where $k=1,2,\dots,2^M$.
To be consistent with our claims we interpret the first element of Eq.[\ref{PS}] which is  $\big\{\big\}$ as $0=\eta_1$.
Any general element of the power set can be interpreted as
\begin{equation}
\big\{\lambda_1,\lambda_2,\cdots,\lambda_j\big\}\Leftrightarrow\lambda_1+\lambda_2+\cdots+\lambda_j=\eta_k,
\end{equation}
where $j\leq M$. Adding each value of $\eta_k$'s to $l^{\prime}_{1}$ and $m^{\prime}_1$ simultaneously to generate a new pair of numbers as shown below
\begin{align}
l^{\prime}_{k}=l^{\prime}_{1}+\eta_k && m^{\prime}_{k}=m^{\prime}_{1}+\eta_k.
\end{align}
The pair $l_k^\prime, m_k^\prime$ ($k=1,2,\dots,2^M$) designate which matrix elements of the composite system density will contribute to the matrix element $\rho^{\prime}_{l,m}$ as under
\begin{equation}
\rho^{\prime}_{l,m}=\sum_{k=1}^{2^M}\rho_{l^\prime_k,m^\prime_k}
\end{equation}
\subsection{A detailed illustration to calculate one matrix element of the reduced density matrix}
\begin{figure}[H]
\begin{center}
 \includegraphics[scale=1.0]{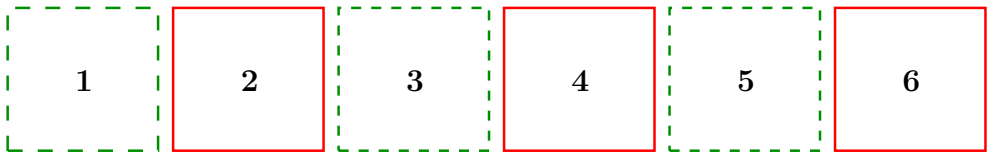}
 \caption{Six qubits quantum system}
 \label{N6}
\end{center}
\end{figure}

We will illustrate the above method by the following example which will be useful for coders, for clarity we study in detail how to calculate one of the matrix elements of the reduced density matrix. However, a more rigorous calculation of computing the total reduced density matrix is relegated to the appendix. Let us suppose we have $6$ spins, that is $N=6$ which means that the composite system density matrix dimension is
$2^6\times2^6$ ($64\times 64$). We partially trace out all the even position qubits out of the composite system. On tracing out even position qubits the value of $M=3$ for a $6$ qubit quantum system. The $6$ qubit quantum system is shown in Fig.[\ref{N6}], the red boxes are for the even positions qubits that are to be traced out and green dashed boxes are for the odd positions qubits.

 The dimension of the reduced density matrix $\rho^{\prime}$
is $2^3\times2^3$. As an illustration let us calculate the value of the matrix element $\rho^{\prime}_{2,1}$ using our method. To do that, first we convert the decimal number $l=2$ to binary 
equivalent number $p$ and $\vert p\rangle$ is one of the vector in the computational basis.

\begin{figure}[h]
\begin{center}
 \includegraphics[scale=2.0]{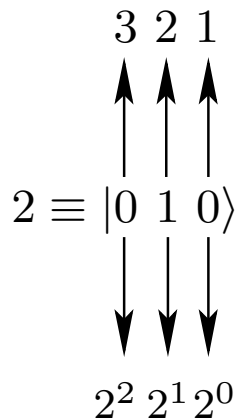}
 \caption{Conversion of $2$ to $\vert p \rangle$}
 \label{bi2}
\end{center}
\end{figure}
In Fig.[\ref{bi2}] the up arrows refers to the positions of the qubits starting from right to left and down arrows refers to the binary place value
of the binary equivalent number. Similarly conversion of decimal number $m=1$ into the binary equivalent number $q$ and $\langle q \vert$  is one
of the vectors in the computational basis of the dual space

\begin{figure}[h]
\begin{center}
 \includegraphics[scale=2.0]{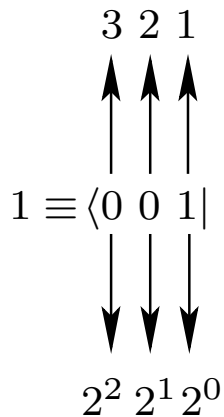}
 \caption{Conversion of $1$ to $\langle q \vert$}
 \label{bi1}
\end{center}
\end{figure}
\begin{figure}[H]
\hspace{2.5cm} \includegraphics[scale=1.15]{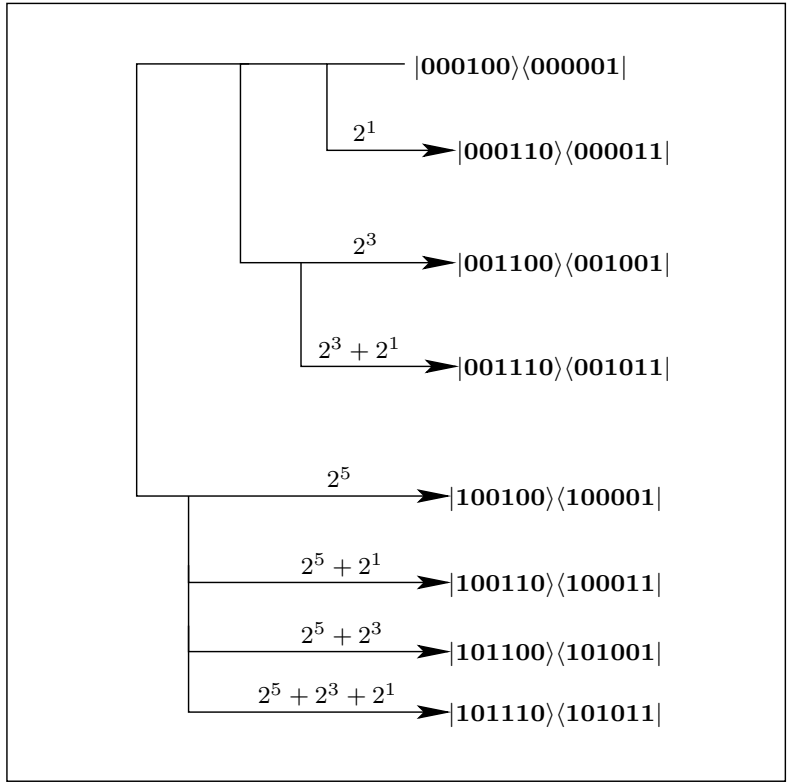}
 \caption{Addition of all the combination to $\vert p^{\prime}\rangle$ and $\langle q^{\prime} \vert$ simultaneously}
 \label{ptr}
\end{figure}

We take another set of two binary numbers $p^{\prime}$ and $q^{\prime}$, which are a string of $6$ numbers and write it in terms of the computational basis vectors as $\vert p^{\prime}\rangle$ and 
$\langle q^{\prime}\vert$. In the even positions starting from right to left of $\vert p^\prime \rangle$ we put zero and leave the odd positions empty for now as indicated by the dashes as shown below
\begin{equation}
 \vert p^{\prime}\rangle=\vert 0~\_~0~\_~0~\_\rangle . \label{pp1}
\end{equation}
Now in the odd position of Eq.[\ref{pp1}] we sequentially putting the values of the $\vert p\rangle$, we get
\begin{equation}
 \vert p^{\prime}\rangle=\vert 000100\rangle . \label{pcm}
\end{equation}
Similarly we can write
\begin{equation}
 \langle q^{\prime}\vert=\langle 000001\vert.
\end{equation}
The binary to decimal conversion of the numbers $p^{\prime}$ and $q^{\prime}$ are $l_1^{\prime}=4$ and $m^{\prime}_1=1$ respectively. As we are partially tracing out even position qubits out of the composite system and the binary place values of these even positions are $\lambda_1=2^1,\lambda_2=2^3,\lambda_3=2^5$. Now we can write the set $S$ as under
\begin{equation}
S=\{\lambda_1,\lambda_2,\lambda_3\}.
\label{s3}
\end{equation}
The power set for the Eq.[\ref{s3}] can be written as
\begin{equation}
P(S)=\bigg\{\big\{\big\},\big\{\lambda_1\big\},\big\{\lambda_2\big\},\big\{\lambda_3\big\},\big\{\lambda_1,\lambda_2\big\},\big\{\lambda_1,\lambda_3\big\},\big\{\lambda_2,\lambda_3\big\},\big\{\lambda_1,\lambda_2,\lambda_3\big\}\bigg\}
\label{psx}
\end{equation}
The total number of $\eta_k$'s element is equal to $2^3$. The $\eta_k$'s element can be written as
$\eta_1=0,\eta_2=\lambda_1,\eta_3=\lambda_2,\eta_4=\lambda_3,\eta_5=\lambda_1+\lambda_2,\eta_6=\lambda_1+\lambda_3,\eta_7=\lambda_2+\lambda_3,\eta_8=\lambda_1+\lambda_2+\lambda_3$. Now add all the $\eta_k$'s element to $l_1^{\prime}=4$ and $m^{\prime}_1=1$ simultaneously as discussed in section \ref{cmpt} 
\begin{align}
l^{\prime}_{1}=4+\eta_1=4 &&  m^{\prime}_{1}=1+\eta_1=1, \label{cm0}\\
 l^{\prime}_{2}=4+\eta_2=6 &&  m^{\prime}_{2}=1+\eta_2=3, \label{cm1}\\
 l^{\prime}_{3}=4+\eta_3=12 &&  m^{\prime}_{3}=1+\eta_3=9, \label{cm2}\\
 l^{\prime}_{4}=4+\eta_4=36 &&  m^{\prime}_{4}=1+\eta_4=33, \label{cm3}\\
 l^{\prime}_{5}=4+\eta_5=14 &&  m^{\prime}_{5}=1+\eta_5=11, \label{cm4}\\
  l^{\prime}_{6}=4+\eta_6=38 &&  m^{\prime}_{6}=1+\eta_6=35, \label{cm5}\\
   l^{\prime}_{7}=4+\eta_7=44 &&  m^{\prime}_{7}=1+\eta_7=41, \label{cm6}\\
    l^{\prime}_{8}=4+\eta_8=46 &&  m^{\prime}_{8}=1+\eta_8=43. \label{cm7}
\end{align}
 As we have discussed earlier that $l^{\prime}_i$ and $m^{\prime}_i$ $(i=1,2\cdots8)$ designates the row and column of the composite system density matrix $\rho$ respectively. The Fig.[\ref{ptr}] explains the procedure what we had written from Eq.[\ref{cm0}] to Eq.[\ref{cm7}] in computational basis representation. To be more explicit, In Fig.[\ref{ptr}] the ket vectors represents the binary conversion of the number $l_i^\prime$ $(i=1,2\cdots8)$ and dimension of the ket vector is $64\times1$. The position $l_i^\prime{}$ in the corresponding ket vector contains the value $1$ and rest all the positions in that ket vector is $0$. Similarly the bra vectors are the binary conversion of the number $m_i^\prime$ $(i=1,2\cdots8)$ and dimension of the bra vector is $1\times64$. The position $m_i^\prime{}$ in the corresponding bra vector contains the value $1$ and rest all the positions in that bra vector is $0$. The coefficients of these outer product matrices are added to give the value of the matrix element $\rho^{\prime}_{2,1}$ of the reduced density matrix. Adding the coefficients that contribute to the matrix element of the reduced density matrix takes computationally less time to calculate than performing matrix multiplication or calculating inner products.
The value of the matrix element $\rho^{\prime}_{2,1}$ is then,
\begin{equation}
 \rho^{\prime}_{2,1}=\rho_{4,1}+\rho_{6,3}+\rho_{12,9}+\rho_{36,33}+\rho_{14,11}+\rho_{38,35}+\rho_{44,41}+\rho_{46,43}.
\end{equation}

\section{Merits and other features of our Method of partial tracing}
\label{finalmerit}
\begin{figure}[h]
\begin{center}
 \includegraphics[height=2cm,width=14cm]{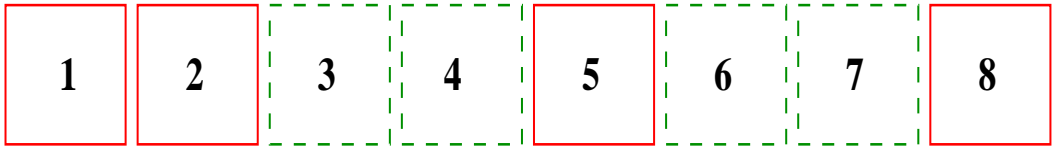}
 \caption{Eight qubits system }
 \label{qbt}
\end{center}
\end{figure}

The method of partial tracing which we have demonstrated can be used not only for a multipartition system but also for a bipartition system. Our method of calculating partial trace is more easier. This fact becomes more evident as we go
to higher dimensional Hilbert spaces which occurs in the finite size spin models of magnetism like Ising model \cite{ising}, Heisenberg model \cite{bethe}, Majumdar-Ghosh model \cite{mg}, AKLT model \cite{aklt} etc. where we have to deal with a large number of qubits (electron which are spin $\frac{1}{2}$ particles). Let us suppose we have an eight qubit system out of which we partially trace
out four qubits. In Fig.[\ref{qbt}] the boxes which are in red are the qubits that are to be partially traced out and the dashed boxes which are in green are the qubits 
which will not be traced out of the system. The eight qubit system is represented by the pure state vector $\vert \psi\rangle$ and its pure state density matrix is
$\rho=\vert\psi\rangle\langle\psi\vert$.

In the method given in \cite{jonas} the algorithm is to partially trace out qubits from the
left hand side till it reaches a qubit in the green dashed box. In the above case the qubits $1$ and $2$ are to be traced out from the left hand side, so using Eq.[\ref{jmpt_bp}] in the bipartition
method of partial tracing it will trace out qubits $1$ and $2$ from the density matrix $\rho$. After tracing out qubits $1$ and $2$ we get the reduced density matrix $\rho^{\prime}$, now from $\rho^{\prime}$ it will partially trace out qubits from the right hand side till it reaches a qubit in green dashed box. In the case under consideration qubit $8$ will be traced out using Eq.[\ref{jmpt_bp}]. Once it traces out qubits from the left and right it uses the Eq.[\ref{jmpt_mp}] to trace out remaining qubits if there 
are any, in the above case qubit $5$ will be traced out and we get the final reduced density matrix 
$\rho^{\prime\prime}$. However, in the method we have given, using the state vector $\vert \psi\rangle$ we can directly generate the final reduced density matrix which in above case is $\rho^{\prime\prime}$ 
without generating any intermediate reduced density matrix like $\rho^{\prime}$. Our method is more efficient in handling larger dimensional density matrices due to its inherent nature of simplicity. For example we have $N=26$ qubits, which means the dimension of the pure state vector $\vert \psi\rangle$ is $2^{26}\times1(2^{26}=67108864)$ (\textit{around a staggering 60 million set of numbers}) and we partially trace out any of the $24$ qubits out of the system, it will give the result within ten seconds in a normal desktop of $4$ GB RAM, Intel i3 processor. However, if we use the method as given in \cite{jonas}, we cannot do this partial tracing in a normal desktop because the dimension of the intermediate reduced density matrix is quite large to handle for a normal desktop. It is also noteworthy to mention that the upper bound on the applicability of the method as given in \cite{jonas} for both pure and mixed state is $N=14$ tracing out $8$ qubits out of the quantum system. Whereas in our method, we reiterate the fact that for a pure state we can maximum go upto $N=26$ tracing out $24$ qubits out of the quantum system. The upper bound on $N$ in a pure state for both methods is different because of the inherent construction of the algorithm. Another important feature of our technique is that it always gives the correct result for any random combination of partial trace but the technique as given in \cite{jonas} is computationally not efficient in achieveing this due to the inherent complexity of the algorithm which has been explained already in this section. The above statement can be further understood from the purview of spin chains that, if we are interested to study an odd or an even part of the spin chain, for instance comb entanglement \cite{comb}, our method gives us an edge over other known methods. It is also worth mentioning that the computations which we have done and presented are done in serial programming. However, if we achieve a successful parallelization of the codes then our method can be made superfast in calculating partial traces.

\section{Summary and conclusions}
Partial trace is a very important mathematical operation for studying the details of a subsystem. Not only that, it is used in the calculation of a variety of entanglement measures. If we use the unoptimized code to do partial tracing it may take considerable amount of time to compute the reduced density matrices and the computational time
increases exponentially with increase in the numbers of qubits. As discussed in this paper, if we use our method for partial tracing then we can handle larger Hilbert spaces within the capacity of our normal desktop. This work may be of importance for the working physicist in the area of quantum information and who often have trouble in dealing with large matrices and quantum states. In fact, we conjecture that, using our method coupled with supercomputers we can considerably reduce the computational time even while dealing with very high dimensional Hilbert spaces. Not only these, the method given by us can calculate partial trace between random partitions of subsystems with great ease.
\section{Acknowledgements}
I would like to thank Miss Payal Solanki (M.Sc. student, VNIT Nagpur) for interesting discussions.
\appendix
\section{A complete example of calculating the reduced density matrix using our method}
In this appendix, we have worked out a complete example of calculating the reduced density matrix by our method. This may help the reader to attain more clarity in understanding the method. Consider a four-qubit quantum system which is represented by the tensor product of two Bell states as,
\begin{equation}
\vert \psi\rangle=\frac{1}{\sqrt{2}}\bigg[\vert 01\rangle_{AB} - \vert 10\rangle_{AB}\bigg]\otimes \frac{1}{\sqrt{2}}\bigg[\vert 01\rangle_{CD} - \vert 10\rangle_{CD}\bigg].
\end{equation}
The pure state density matrix is written as.
\begin{eqnarray}
\rho& =&\vert \psi\rangle\langle \psi\vert \nonumber \\
&=& \frac{1}{4} \bigg[\vert 0101\rangle_{ABCD} -\vert 0110\rangle_{ABCD} -\vert 1001\rangle_{ABCD} +\vert 10101\rangle_{ABCD} \bigg] \nonumber \\ &&  \bigg[{}_{ABCD}\langle 0101\vert-{}_{ABCD}\langle 0110\vert - {}_{ABCD}\langle 1001\vert +{}_{ABCD}\langle 10101\vert \bigg].
\label{dens}
\end{eqnarray}
Partially tracing out subsystems $B$, $C$ and $D$ out of the density matrix $\rho$, the qubit positions of the subsystem are $3$, $2$  and $1$ respectively and the binary place values of the qubit positions are $\lambda_3=2^2$, $\lambda_2=2^1$ and $\lambda_1=2^0$ respectively. The set $S$ can be written as
\begin{equation}
S=\{\lambda_1,\lambda_2,\lambda_3\}.
\end{equation}
The power set $P(S)$ can be written same as Eq.[\ref{psx}] and the $\eta_k$'s values are written as $\eta_1=0,\eta_2=\lambda_1,\eta_3=\lambda_2,\eta_4=\lambda_3,\eta_5=\lambda_1+\lambda_2,\eta_6=\lambda_1+\lambda_3,\eta_7=\lambda_2+\lambda_3$ and $\eta_8=\lambda_1+\lambda_2+\lambda_3$.
Computing the matrix element of the reduced density matrix $\rho^\prime_{00}$, the decimal numbers $l$ and $m$ are $0$ and $0$. Converting these decimal numbers into binary numbers $p$ and $q$ of one bit string, we get $0$ and $0$. Now calculating the binary number $p^\prime$ a string of four bits, starting from right to left the qubit positions $3$, $2$  and $1$  will be containing zero
\begin{equation}
p^\prime=\_000.
\end{equation}  
The qubit position $4$ will be filled by the number in the binary number $p$.
\begin{equation}
p^\prime=0000. \label{pval}
\end{equation} 
Converting back the binary number $p^\prime$ into decimal number $l_1^\prime=0$ and similarly we found out $m_1^\prime=0$. Now adding the $\eta_k$'s values to $l_1^\prime$ and $m_1^\prime$ simultaneously, we get 
\begin{align}
l^{\prime}_{1}=0+\eta_1=0 &&  m^{\prime}_{1}=0+\eta_1=0, \label{den0} \\
 l^{\prime}_{2}=0+\eta_2=1 &&  m^{\prime}_{2}=0+\eta_2=1, \label{den1} \\
  l^{\prime}_{3}=0+\eta_3=2 &&  m^{\prime}_{3}=0+\eta_3=2, \label{den2}\\
   l^{\prime}_{4}=0+\eta_4=4 &&  m^{\prime}_{4}=0+\eta_4=4, \label{den3}\\
    l^{\prime}_{5}=0+\eta_5=3 &&  m^{\prime}_{5}=0+\eta_5=3, \label{den4}\\
    l^{\prime}_{6}=0+\eta_6=5 &&  m^{\prime}_{6}=0+\eta_6=5, \label{den5}\\
    l^{\prime}_{7}=0+\eta_7=6 &&  m^{\prime}_{7}=0+\eta_7=6, \label{den6}\\
    l^{\prime}_{8}=0+\eta_8=7 &&  m^{\prime}_{8}=0+\eta_8=7. \label{den7}
\end{align}
The matrix element $\rho_{00}^\prime$ can be written as
\begin{eqnarray}
\rho_{00}^\prime &=& \rho_{l_1^\prime m_1^\prime}+\rho_{l_2^\prime m_2^\prime}+\rho_{l_3^\prime m_3^\prime}+\rho_{l_4^\prime m_4^\prime}+\rho_{l_5^\prime m_5^\prime}+\rho_{l_6^\prime m_6^\prime}+\rho_{l_7^\prime m_7^\prime}+\rho_{l_8^\prime m_8^\prime},\\
\rho_{00}^\prime &=& \rho_{0,0}+\rho_{1,1}+\rho_{2,2}+\rho_{3,3}+\rho_{4,4}+\rho_{5,5}+\rho_{6,6}+\rho_{7,7}.
\end{eqnarray}
Using Eq.[\ref{dens}] in the above equation, we get
\begin{eqnarray}
\rho_{00}^\prime &=& 0+0+0+0+0+0.25+0.25+0,\\
\rho_{00}^\prime &=& 0.5. \label{den4_1}
\end{eqnarray}
We now calculate the matrix element $\rho^\prime_{01}$, the decimal numbers $l$ and $m$ are $0$ and $1$. On converting decimal numbers $l$ and $m$ into one bit string on binary numbers $p=0$ and $q=1$ respectively. The binary number $p^\prime$ can be written as
\begin{equation}
p^\prime=0000.
\end{equation}
But while calculating binary number $q^\prime$ the qubit position starting from right to left $1$, $2$ and $3$ contains zero.
\begin{equation}
q^\prime=\_000.
\end{equation}
Now insert value of binary number $q$ into the $4^{th}$ bit position and we get
\begin{equation}
q^\prime=1000.
\end{equation}
The binary to decimal conversion of the numbers $p^\prime$ and $q^\prime$ are $l_1^\prime=0$ and $m_1^\prime=8$. Now by replacing value of $l_1^\prime=0$ and $m_1^\prime=8$ from eqn.(\ref{den0}) to eqn.(\ref{den7}) we get the new values of $l_2^\prime,m_2^\prime,l_3^\prime\cdots l_8^\prime,m_8^\prime$. So the matrix element $\rho^\prime_{01}$ can be written as
\begin{eqnarray}
\rho^\prime_{01} &=& \rho_{0,8}+\rho_{1,9}+\rho_{2,10}+\rho_{3,11}+\rho_{4,12}+\rho_{5,13}+\rho_{6,14}+\rho_{7,15},\\
\rho^\prime_{01} &=& 0+0+0+0+0+0+0+0,\\
\rho^\prime_{01} &=& 0. \label{den4_2}
\end{eqnarray}
From eqn.(\ref{den4_2}) we can say that $\rho^\prime_{10}=0$ because $\rho^\prime$ is a Hermitian matrix and is symmetric. We follow the similar procedure to calculate the matrix element $\rho^\prime_{11}$.
\begin{eqnarray}
\rho^\prime_{11} &=& \rho_{8,8}+\rho_{9,9}+\rho_{10,10}+\rho_{11,11}+\rho_{12,12}+\rho_{13,13}+\rho_{15,14}+\rho_{15,15},\\
\rho^\prime_{11} &=& 0+0.25+0.25+0+0+0+0+0,\\
\rho^\prime_{11} &=& 0.5. \label{den4_3}
\end{eqnarray}
Using eqn.(\ref{den4_1}), (\ref{den4_2}) and (\ref{den4_3}) we can write the matrix $\rho^\prime$ as
\begin{equation}
\rho^\prime=\left(\begin{array}{cc}
0.5 & 0\\
0 & 0.5
\end{array}\right)
\end{equation}

\end{document}